\documentclass[twocolumn,pra,amsmath,amssymb,showpacs,aps]{revtex4-1}
\usepackage{graphicx}
\begin{document}

\newcommand{\ket}[1]    {\left| #1 \right\rangle}
\newcommand{\av}[1]    {\langle #1 \rangle}

\title{Simple atom interferometer in a double-well potential}

\author{Karol Gietka}
\author{Jan Chwede\'nczuk}
\affiliation{
  Faculty of Physics, University of Warsaw, ul. Ho\.za 69, 00-681 Warsaw, Poland\\
}
\begin{abstract} 
  We present a detailed study of an atom interferometer which can be realized in a double-well potential. 
  We assume that the interferometric phase is imprinted in the presence of coherent tunneling between
  the wells. We calculate the ultimate bounds for the estimation sensitivity and show how they relate to the precision of the Mach-Zehnder interferometer. The interferometer presented here
  allows for sub shot-noise sensitivity when fed with the spin-squeezed states with reduced either the relative population imbalance or the relative phase. We also calculate the precision
  of the estimation from the population imbalance and show that it overcomes the shot-noise level when the entangled squeezed-states are used at the input. 
\end{abstract}

\pacs{
  37.25.+k,
  03.75.-b,
  03.75.Lm
}

\maketitle

\section{Introduction} 
The key objective of quantum interferometry is to enhance the estimation precision $\Delta\theta$ of an unknown parameter $\theta$ using non-classical correlations
as a resource.
The reference value $\Delta \theta_{\rm SN} = \frac1{\sqrt m}\frac1{\sqrt N}$ is called the shot-noise limit (SNL), 
with $N$ being the number of particles passing through the interferometer and $m$ being the number of measurement repetitions. 
The SNL is the best achievable sensitivity in the classical two-mode interferometry.
Only in the presence of useful particle entanglement, the SNL can be surpassed \cite{giovannetti2004quantum,pezze2009entanglement} to give $\Delta\theta<\Delta\theta_{\rm SN}$. 
Therefore, quantum interferometry can be viewed from two perspectives. From one point of view, the stress is put on the preparation of a usefully entangled quantum state which 
together with the estimation protocol cooperate to give $\Delta\theta<\Delta\theta_{\rm SN}$. 
From the other point of view, interferometry is a tool for detecting quantum correlations in many-body systems. 
In this case, the value of $\Delta\theta$ is treated as a probe of the particle entanglement. 

The $\Delta\theta$ can be evaluated using the Cramer-Rao lower bound (CRLB) \cite{holevo2011probabilistic}. This important theorem links the sensitivity with the Fisher information
\begin{equation}\label{crlb}
  \Delta\theta\geqslant\frac1{\sqrt m}\frac1{\sqrt F}.
\end{equation}
The value of $F$ depends on all the three steps of the interferometric sequence: the preparation of the state which enters the device, 
the type of the interferometric transformation, and the measurement performed at the output to obtain $\theta$. According to the definition of the SNL, $F>N$ signals the particle entanglement 
\cite{pezze2009entanglement}. 
However, in most experimental situations, it is very difficult to directly measure the value of $F$. 
The solution to this problem is to replace the Fisher information in Eq.~(\ref{crlb}) with some other physical quantity which is 
more accessible in the laboratory. However, this new quantity sets a weaker constraint than the CRLB~(\ref{crlb}). 

This approach is illustrated by a broad use of the spin-squeezing parameter $\xi_n^2$ \cite{kitagawa1993squeezed,wineland1994squeezed}. 
It is proportional to the fluctuations of the number of particles between the two modes divided by the visibility of the one-body fringes. 
Spin-squeezed states ($\xi_n^2<1$) are particle-entangled and potentially useful for quantum metrology. Recently, the spin-squeezing 
has been generated in two-mode quantum systems \cite{esteve2008squeezing,appel2009mesoscopic,gross2010nonlinear,riedel2010atom,leroux2010orientation,chen2011conditional,berrada2013integrated}. 
A similar technique to detect the non-classical correlations was used in a collection of atoms scattered from a single Bose-Einstein condensate in the spin-changing collisions \cite{lucke2011twin}. 

A usefully entangled quantum state passes through a metrological device, for example, the Mach-Zehnder interferometer (MZI) which is realized in three steps. 
First, the two-mode state goes through a beam splitter, then a  phase $\theta$ is imprinted on one of the arms, and finally another beam splitter mixes the modes to yield an interferometric signal. 
The MZI can benefit from the quantum correlations present in the spin-squeezed state to provide the sensitivity $\Delta\theta$ below the SNL
\cite{pezze2009entanglement,pezze_mzi}. 
Another type of an interferometric sequence is based on the Bloch oscillations of a gas in a double- (or many-) 
well potential \cite{ferrari,poli,fattori,clade2006determination,carusotto2005sensitive,battesti2004bloch,dahan1996bloch, clade2014bloch}. In this scenario, the external force drives the coherent oscillations
between the sites of the periodic potential. Therefore, in contrast to the MZI, the mode mixing occurs simultaneously with the phase imprint.

In this work, we study in detail a performance of an interferometer where the phase imprint is accompanied by the tunneling of the gas between the two sites of the trapping potential. 
In Section \ref{sec_model}, we introduce a simple model for the two-mode system of ultra-cold bosons trapped in a double-well potential. 
We determine the evolution operator and present the family of input states convenient for our analysis. In Section \ref{sec_qfi}, using the notion of the quantum Fisher information, 
we calculate the ultimate bounds for the precision of such a double-well interferometer. In Section \ref{sec_est}, we calculate the precision for a particular choice of the estimation protocol
and compare these results to the ultimate bounds. The conclusions are contained in Section \ref{sec_con}. This work is an extension of a previous study \cite{chwedenczuk2010rabi} where the outline of
the theory of such an interferometer was presented.

\section{The model}
\label{sec_model}

We consider a collection of $N$ non-interacting bosons trapped in a symmetric double-well potential $V_{\rm dw}(x)$. The system is driven into the oscillations between
the two wells due to the presence of an external force with a potential $V(x)$.
The objective of the following inquiry is to examine how, and with what precision, the strength of $V(x)$ can be determined. 
To accomplish this task, we employ the two-mode approximation where the field operator reads
\begin{equation}\label{op}
  \hat\Psi(x)=\psi_a(x)\hat a+\psi_b(x)\hat b.
\end{equation}
Here $\hat a$/$\hat b$ annihilates a boson in a left/right potential well, and $\psi_{a/b}(x)$ is a corresponding spatial wave-packet. 
The Hamiltonian of the system is
\begin{equation}
  \hat H=\int\!\! \mathrm{d}x\,\hat\Psi^\dagger(x)\left[-\frac{\hbar^2}{2M}\frac{\partial^2}{\partial x^2}+V_{\rm dw}(x)+V(x)\right]\hat\Psi(x)\label{ham1},
\end{equation}
where $M$ is the atomic mass. We employ the definition of the Josephson energy $E_J$ and the detuning $\delta$, i.e.,
\begin{subequations}
  \begin{eqnarray}
    &&E_J=2 \int\! \mathrm{d}x\,\psi^*_{a}(x)\left[-\frac{\hbar^2}{2M}\frac{\partial^2}{\partial x^2}+V_{\rm dw}(x)\right]\psi_b(x)\\
    &&\delta=\int \mathrm{d}x\left(|\psi_a(x)|^2-|\psi_b(x)|^2\right)V(x)
  \end{eqnarray}
\end{subequations}
to obtain that, up to the constant terms, the Hamiltonian (\ref{ham1}) can be expressed in a compact form
\begin{equation}
  \hat H=-E_J\hat J_x+\delta\hat J_z.\label{ham}
\end{equation}
The $\hat J_x$ and $\hat J_z$ angular momentum operators which appear above, together with the $y$-component, read
\begin{subequations}
  \begin{eqnarray}
    &&\hat J_x=\frac12(\hat a^\dagger\hat b+\hat a\hat b^\dagger)\\
    &&\hat J_y=\frac1{2i}(\hat a^\dagger\hat b-\hat a\hat b^\dagger)\\
    &&\hat J_z=\frac12(\hat a^\dagger\hat a-\hat b^\dagger\hat b).\label{Jz}
  \end{eqnarray}
\end{subequations}
These operators form a Lie algebra $[\hat J_k,\hat J_l]=i\varepsilon_{klm}\hat J_m$.
The Hamiltonian (\ref{ham}) generates the unitary evolution
\begin{equation}\label{evo}
  \hat U=\exp\left[i\varphi(\hat J_x-\epsilon\hat J_z)\right].
\end{equation}
Here $\epsilon = \delta/E_J$ is the ratio of the detuning to the Josephson energy, while $\varphi=E_Jt/\hbar$ is the phase acquired through bare Josephson oscillations. 

To simplify the further analysis, we assume that the initial state which undergoes the evolution (\ref{evo}) is pure
\begin{equation}\label{state}
  \ket\psi=\sum\limits_{n=0}^NC_n|n,N-n\rangle\ \ \ \ \ \mathrm{with}\ \ \ \ \ \sum\limits_{n=0}^N|C_n|^2=1.
\end{equation}
Depending on the coefficients $C_n$,  $\ket\psi$ is either separable or entangled.
Since this initial state is prepared in the absence of the perturbing potential $V(x)$, it is
reasonable to assume that it is path-symmetric, i.e., $C_n=C_{N-n}$. This symmetry vastly simplifies the following discussion through the set of algebraic relations
\begin{equation}\label{path}
  \av{\hat J_y}=\av{\hat J_z}=\av{\hat J_x\hat J_y}=\av{\hat J_x\hat J_z}=\av{\hat J_y\hat J_z}=0.
\end{equation}

The Hamiltonian (\ref{ham}) leads to various types of interferometric schemes depending on the ratio of the Josephson energy to the detuning $\delta$.
One limiting case is when tunneling is fully suppressed during the action of the external force, i.e., $\epsilon\rightarrow\infty$. 
In such a case, the interferometric transformation consists of a bare phase-imprint because the evolution operator (\ref{evo}) simplifies to
\begin{equation}\label{imp}
  \hat U_{\rm ph}=e^{-i\theta\hat J_z},
\end{equation}
where $\theta=\delta t/\hbar$. To obtain some $\theta$-dependent signal, additional mode-mixing manipulation is necessary. Usually, two distinct scenarios are considered to accomplish
this task. In the first one, the phase imprint (\ref{imp}) is preceded and followed by a pair of beam-splitters, and the full cycle is the MZI with
an effective evolution operator
\begin{equation}\label{mzi}
  \hat U_{\rm MZI}=e^{-i\theta\hat J_y}.
\end{equation}
Note that when the two modes represent atomic internal degrees of freedom, the beam-splitters can be realized by applying a precisely crafted rf-pulse 
\cite{gross2010nonlinear,lucke2011twin,riedel2010atom}. However, when the modes are spatially separated, as in a double-well potential 
\cite{gati2006noise,jo2007long,schumm2005matter,bohi2009coherent,berrada2013integrated}, the beam-splitter is more difficult to implement. 
In an alternative scenario of obtaining the interferometric signal from the evolution (\ref{imp}), the gas is simply released from the trap. In the far-field regime, an interference
pattern is formed, and $\theta$ can be inferred from the measurements of positions of individual atoms \cite{chwed_njp}, for instance, from a least-square fit
of the one-body density to the acquired data. In such a case, the sub shot-noise (SSN) sensitivity can be achieved with the phase-squeezed states \cite{grond2010atom,chwed_njp}. 
However, to reach the Heisenberg scaling,
the knowledge of the full $N$-body correlations is necessary \cite{chwedenczuk2011phase}, which for large $N$ is practically impossible.

As underlined in the Introduction, we will analyze the interferometer performance when both the tunneling and the detuning compete at the same time. 
Formally, this means that $\epsilon\lesssim1$ and the evolution operator is
given by the full expression (\ref{evo}) rather than the simplified (\ref{imp}). This type of evolution has one clear advantage over the above scenario. Namely,
the modes are mixed already during the interaction of the gas with the external field, and no addition to the interferometric sequence is necessary.

It is worth to note that the Hamiltonian (\ref{ham}) generates the rotation of the composite spin-$\frac N2$ vector on the Bloch sphere. For such a transformation, 
states which give high metrological precision are those which have reduced fluctuations in the direction orthogonal to the rotation. For instance, if the interferometer rotates the state
around the $y$-axis---as in the MZI (\ref{mzi})---the useful entanglement is related to the spin-squeezing in the $z$-direction. It might seem that finding a usefully entangled state 
for the Hamiltonian (\ref{ham}) should be easy---one should just squeeze the state in a direction orthogonal to the vector with the Cartesian coordinates $(-E_J,0,\delta)$.
However, the knowledge of the direction of this vector is equivalent to the knowledge of $\delta$ which, actually, is the parameter to be estimated. Although some adaptive methods could be used to first roughly estimate $\delta$ and
then prepare the properly entangled states, we assume that $\delta$ remains completely unknown and the input states are typical for the two-mode atom interferometry.

Finally, note that during the evolution governed by the Hamiltonian (\ref{ham}), the two-body interactions are absent. This can be achieved by tuning the 
scattering length using the Feschbach resonances \cite{fesch,pethick}. Although our analysis assumes a complete lack of
interactions, some residual two-body collisions might be present \cite{fattori}. In a more realistic model, they should be included either perturbatively in the analytical calculation or numerically.

\section{Ultimate precision -- quantum Fisher information}
\label{sec_qfi}

In the first step, we calculate the maximal attainable precision of the estimation of $\delta$. 
With this result at hand, we will have a possibility to judge the efficiency of a simple estimation protocol. 
Note that usually the interferometer is characterized by its phase sensitivity $\Delta\theta$. Here we use $\Delta\delta$, which is the precision of the estimation of the
sole parameter $\delta$. The phase sensitivity can be retrieved through a multiplication of $\Delta\delta$ by $t/\hbar$, where $t$ is the time span of the interferometric sequence.

The ultimate precision $\Delta\delta$, which is optimized over all the estimation strategies, is determined by the quantum Fisher information (QFI) denoted by $F_Q$. 
Its value depends on the input state $\ket\psi$ and the Hamiltonian (\ref{ham}) which introduces
the $\delta$-dependence into the system. For pure states, the ultimate CRLB is  \cite{braunstein1994statistical}
\begin{equation}\label{qfi}
  \Delta\delta\geqslant\frac1{\sqrt m}\frac1{\sqrt{F_Q}}=\frac1{\sqrt m}\frac1{\sqrt{4\av{(\Delta\hat h)^2}}}.
\end{equation}
The variance $\av{(\Delta\hat h)^2}=\av{\hat h^2}-\av{\hat h}^2$ is calculated for the initial state $\ket\psi$, and $\hat h$ is an operator which generates the transformation
\begin{equation}
  i\partial_\delta\ket{\psi(\delta)}=\hat h\ket{\psi(\delta)}.
\end{equation}
Using $\ket{\psi(\delta)}=\hat U\ket\psi$, we obtain that $\hat h$ is related to the evolution operator (\ref{evo}) by the expression
\begin{equation}\label{gen}
  \hat h=i\frac{\partial\hat U}{\partial\delta}\hat U^\dagger.
\end{equation}
Note that it is convenient to express the sensitivity (\ref{qfi}) in units of $\delta$, i.e., to replace $\hat h$ with $\delta\cdot\hat h$.
Calculation of the QFI using Equations (\ref{evo}), (\ref{qfi}), and (\ref{gen}) is straightforward. The commutation relations of the angular momentum operators give the rescaled generator equal to
\begin{equation}\label{h}
  \hat{h}=h_x\hat J_x+h_y\hat J_y+h_z\hat J_z,
\end{equation}
where the three coefficients $h_x$, $h_y$, and $h_z$ read
\begin{subequations}\label{coeff}
  \begin{eqnarray}
    &&h_x=\frac{\epsilon^2}{\epsilon ^2+1}\left(\frac{\sin \left(\varphi  \sqrt{\epsilon ^2+1}\right)}{\sqrt{\epsilon ^2+1}}-\varphi\right)\\
    &&h_y=\frac{\epsilon}{\epsilon ^2+1}\left(1-\cos \left(\varphi  \sqrt{\epsilon ^2+1}\right)\right)\\
    &&h_z=\frac{\epsilon^3}{\epsilon ^2+1}\left(\frac{\sin\left(\varphi\sqrt{\epsilon^2+1}\right)}{\epsilon^2\sqrt{\epsilon^2+1}}+\varphi\right).
  \end{eqnarray}
\end{subequations}
Substituting Eq.~(\ref{h}) into Eq.~(\ref{qfi}), we obtain for
the path-symmetric states (\ref{path})
\begin{equation}\label{qfi2}
  F_Q=4\left(h_x^2\av{(\Delta\hat J_x)^2}+h_y^2\av{\hat J_y^2}+h_z^2\av{\hat J_z^2}\right).
\end{equation}
Clearly, the QFI is a complicated function of the independent parameters
$\epsilon$ and $\varphi$, and the input state (\ref{state}) by means of 
the two lowest moments of the angular momentum operators.

\begin{figure}[htb!]
  \includegraphics[clip, scale=1]{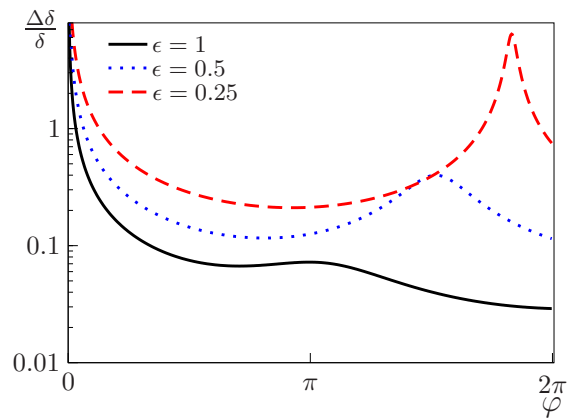}
  \caption{
    (color online) The sensitivity $\Delta\delta$ in units of $\delta$ for a spin-coherent state with $N=100$ 
    plotted as a function of $\varphi$ for three different values of $\epsilon=1$ (solid black line), $\epsilon=0.5$ (dotted blue line), and $\epsilon=0.25$ (dashed red line). 
  }\label{qfi_coh}
\end{figure}

We perform the systematic analysis of Eq.~(\ref{qfi2}) by first fixing the input state---i.e., fixing $\av{(\Delta\hat J_x)^2}$, $\av{\hat J_y^2}$, and $\av{\hat J_z^2}$---and then plotting the QFI as a function
of the other parameters. First, we consider a spin-coherent state 
\begin{equation}\label{scs}
  \ket\psi=\frac1{\sqrt {N!}}\left(\frac{\hat a^\dagger+\hat b^\dagger}{\sqrt 2}\right)^N\ket0,
\end{equation}
which gives $\av{(\Delta\hat J_x)^2}=0$ and $\av{\hat J_y^2}=\av{\hat J_z^2}=\frac N4$. In such a case, the QFI scales linearly with the number of particles (shot-noise scaling), and the sensitivity reads
\begin{equation}
  \frac{\Delta\delta}{\delta}\geqslant\frac1{\sqrt m}\frac1{\sqrt N}\frac1{\sqrt{h_y^2+h_z^2}}.
\end{equation}
We plot this expression in Fig.~\ref{qfi_coh} as a function of $\varphi$ for three different values of $\epsilon$. For small $\epsilon=0.25$, when the tunneling dominates over the detuning,
oscillations are clearly visible. When $\epsilon$ grows, the period of oscillations drops according to Eq.~(\ref{coeff}), and the sensitivity clearly improves with time. This is the result of the
increasing domination of the $\delta\hat J_z$ term in the Hamiltonian (\ref{ham}). 

\begin{figure}[htb!]
  \includegraphics[clip, scale=1]{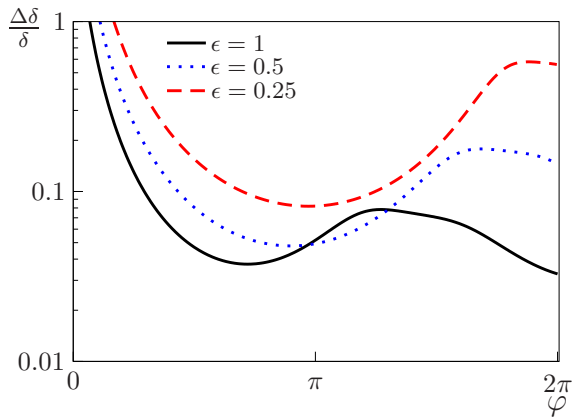}
  \caption{
    (color online) The sensitivity $\Delta\delta$ in units of $\delta$ for a spin-squeezed state of $N=100$ particles with $\xi_n^2=0.15$ 
    plotted as a function of $\varphi$ for three different values of $\epsilon=1$ (solid black line), $\epsilon=0.5$ (dotted blue line), and $\epsilon=0.25$ (dashed red line). 
  }\label{qfi_ns}
\end{figure}

In the next step, we replace the spin coherent state with a spin-squeezed state which has reduced fluctuations of the relative atom number between the two modes
\cite{esteve2008squeezing,appel2009mesoscopic,gross2010nonlinear,riedel2010atom,leroux2010orientation,chen2011conditional,berrada2013integrated}. 
Such a state is characterized with the spin-squeezing parameter \cite{kitagawa1993squeezed,wineland1994squeezed}
\begin{equation}
  \xi_n^2=N\frac{\av{\hat J_z^2}}{\av{\hat J_x}^2}.
\end{equation}
We numerically generate an entangled spin-squeezed state by finding the ground state of the Bose-Hubbard Hamiltonian
\begin{equation}\label{bh}
  \hat H_{\rm bh}=-\hat J_x+\frac\alpha N\hat J_z^2.
\end{equation}
with $N=100$ particles and $\alpha>0$. We take such $\alpha$ to obtain a realistic value $\xi_n^2=0.15$. With this state, we calculate all the moments of the angular momentum operators (\ref{qfi2}) which 
determine the sensitivity (\ref{qfi}). 
In Fig.~\ref{qfi_ns}, we plot the resulting sensitivity in units of $\delta$ as a function of $\varphi$ for the same three values of $\epsilon$ as in Fig.~\ref{qfi_coh}.

\begin{figure}[htb!]
  \includegraphics[clip, scale=1]{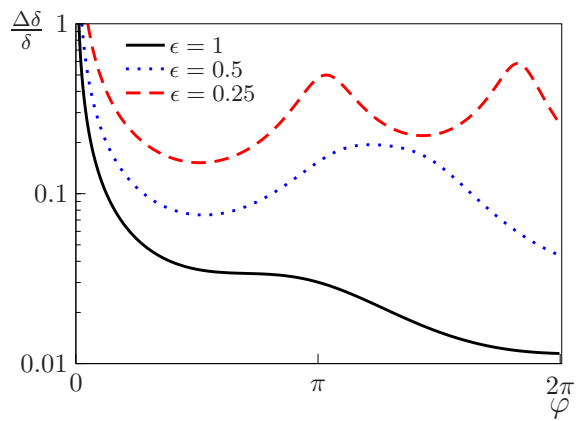}
  \caption{
    (color online) The sensitivity $\Delta\delta$ in units of $\delta$ for a spin-squeezed state of $N=100$ particles with $\xi_\phi^2=0.15$ 
    plotted as a function of $\varphi$ for three different values of $\epsilon=1$ (solid black line), $\epsilon=0.5$ (dotted blue line), and $\epsilon=0.25$ (dashed red line). 
  }\label{qfi_phs}
\end{figure}

Finally, we take a phase-squeezed state, characterized by the following squeezing parameter \cite{grond2010atom,chwed_njp}
\begin{equation}
  \xi_\phi^2=N\frac{\av{\hat J_y^2}}{\av{\hat J_x}^2},
\end{equation}
which we generate with the same Hamiltonian but with $\alpha<0$. We take symmetrically $\xi_\phi^2=0.15$ for $N=100$ particles and plot the analogical sensitivity in Fig.~\ref{qfi_phs}.

We now discuss and compare the results presented in these three figures. First, note that for large $\epsilon$ the phase-squeezed states ($\xi_\phi^2<1$) give better precision than
the number-squeezed ($\xi_n^2<1$). This is because in this regime
the $\delta\hat J_z$ term dominates in the Hamiltonian (\ref{ham}). For the phase-squeezed states, the terms $\av{\hat J_z^2}$ in the QFI dominates over the other two parts, and the coefficient $h_z$ grows
with $\epsilon$. On the other hand, for the number-squeezed states, the $\av{\hat J_y^2}$ dominates over the other parts of the QFI. 
Moreover, according to Equations (\ref{coeff}), this term becomes more important for small $\epsilon$, but we still do not observe a significant improvement of the sensitivity 
between the results for the coherent state from Fig.~\ref{qfi_coh} and the number-squeezed from Fig.~\ref{qfi_ns}. 
To explain this behavior, we expand the coefficients (\ref{coeff}) in a limit of $\epsilon\ll1$ and short times ($\varphi\simeq1$), and we obtain that
\begin{subequations}\label{coeff2}
  \begin{eqnarray}
    &&h_x\simeq\epsilon^2\left(\sin\varphi-\varphi\right)\label{hx2}\\
    &&h_y\simeq\epsilon\left(1-\cos\varphi\right)\\
    &&h_z\simeq\epsilon\sin\varphi.
  \end{eqnarray}
\end{subequations}
Apart from the vicinity of $\varphi=2\pi$,
the $h_x$ coefficient from line (\ref{hx2}) can be neglected compared to $h_y$ and $h_z$, and the lower bound for the sensitivity reads
\begin{equation}\label{jos}
  \frac{\Delta\delta}{\delta}\geqslant\frac1{\sqrt m}\frac1{2\epsilon}\frac{1}{\sqrt{\left(1-\cos \varphi\right)^2\av{\hat J_y^2}+\sin^2\varphi\av{\hat J_z^2}}}.
\end{equation}
Clearly, there is a particular point $\varphi=\pi$ when
\begin{equation}\label{like_mzi}
  \frac{\Delta\delta}\delta\geqslant\frac1{\sqrt m}\frac1{2\epsilon}\frac{1}{\sqrt{4\av{\hat J_y^2}}}.
\end{equation}
This sensitivity closely resembles the ultimate bound for the MZI interferometer (\ref{mzi}), which for pure states reads
\begin{equation}\label{sens_mzi}
  \frac{\Delta\delta_{\rm MZI}}\delta\geqslant\frac1{\sqrt m}\frac1{\theta}\frac{1}{\sqrt{4\av{\hat J_y^2}}}.
\end{equation}
However, since $\theta=\epsilon\times\varphi$, the MZI bound for the sensitivity is $\varphi/2$ times better then Eq.~(\ref{like_mzi}).
This means that the precision (\ref{sens_mzi}), in contrary to (\ref{like_mzi}), improves over time. Nevertheless, since the expansion (\ref{coeff2}) is valid for short times, 
the gain from the time-scaling of Eq.~(\ref{sens_mzi}) over (\ref{like_mzi}) is of the order of  $\pi$. 
Note also that the expression (\ref{like_mzi}) improves for the spin-squeezed states with $\xi_n^2<1$ because for such states $4\av{\hat J_y^2}>N$, but,
on the other hand, it deteriorates when $\epsilon$ drops. These two effect more or less cancel each other for the parameters used in Fig.~\ref{qfi_ns}. However, for large $N$, the improvement
coming from the quantum correlations dominates over the loss of signal, leading to the SSN scaling of the sensitivity.

Another distinguished time is when $\phi=\frac\pi2$. Then, the expression (\ref{jos}) simplifies to
\begin{equation}\label{like_int}
  \frac{\Delta\delta}{\delta}\geqslant\frac1{\sqrt m}\frac1{\epsilon}\frac{1}{\sqrt{4\av{\hat J_y^2}+4\av{\hat J_z^2}}}.
\end{equation}
Interestingly, in this case the sensitivity can be improved over the shot-noise scaling both for the phase-squeezed states, which give $4\av{\hat J_z^2}>N$, or the number-squeezed states, 
which give $4\av{\hat J_y^2}>N$. Still, the loss of the signal for small $\epsilon$ can overshadow the SSN scaling, if $N$ is not sufficiently large.

Finally, we focus on the long-time behavior of the QFI. When $\varphi\gg1$, Equations (\ref{coeff}) simplify and give a following bound for the sensitivity
\begin{equation}\label{long_qfi}
  \frac{\Delta\delta}\delta\geqslant\frac1{\sqrt m}\frac{\epsilon^2+1}{\theta\epsilon}\frac1{\sqrt{4\av{(\Delta\hat J_x)^2}+4\av{\hat J_z^2}\epsilon^2}}.
\end{equation}
If $\epsilon\ll1$ and the state is spin-squeezed with $\xi^2_n<1$, the $4\av{\hat J_z^2}\epsilon^2$ can be safely neglected, and we obtain
\begin{equation}\label{time_x}
  \frac{\Delta\delta}\delta\geqslant\frac1{\sqrt m}\frac{1}{\theta\epsilon}\frac1{\sqrt{4\av{(\Delta\hat J_x)^2}}}.
\end{equation}
If $\ket\psi$ is strongly squeezed, i.e., close to the twin-Fock state $\ket\psi\simeq\ket{\frac N2,\frac N2}$, then $\av{(\Delta\hat J_x)^2}\simeq\av{\hat J_y^2}$ and Equations (\ref{sens_mzi})
and (\ref{time_x}) differ only by a presence of $\epsilon$ in the denominator of the latter. Still, both expressions share the same scaling of the sensitivity with time.
When $\epsilon\simeq1$ and $\ket\psi$ is close to the coherent spin state or is phase-squeezed, then (\ref{long_qfi}) is approximately
\begin{equation}
  \frac{\Delta\delta}\delta\geqslant\frac1{\sqrt m}\frac{\epsilon^2+1}{\theta\epsilon^2}\frac1{\sqrt{4\av{\hat J_z^2}}}.
\end{equation}
This sensitivity breaks the SNL, scales inversely in time, and is only $\frac{\epsilon^2+1}{\epsilon^2}\simeq2$ times worse than the ultimate bound for the pure phase-imprint (\ref{imp}).

To summarize this Section, we have calculated the ultimate bound for the sensitivity of the double-well interferometer. We have shown, that it betrays the characteristic oscillatory behavior due
to the presence of the Josephson term in the Hamiltonian (\ref{ham}). We have also shown that for some particular instants of time, the QFI can be improved beyond the SNL with either
the number-squeezed or the phase-squeezed states. At long times and with spin-squeezed ($\xi_n^2<1$) input states, the sensitivity closely resembles the precision of the MZI, whereas with phase-squeezed
states ($\xi_\phi^2<1$), it is almost as good as for a pure phase-imprint.

\section{Estimation from the population imbalance}
\label{sec_est}

We now focus on a particular scheme of estimation based on the measurement of the population imbalance.
The sequence we consider is following. First, the input state (\ref{state}) evolves according to Eq.~(\ref{evo}).
Next, a population imbalance $n$ between the two sites is measured. If this data is used to estimate the value of $\delta$, the CRLB reads
\begin{equation}\label{crlb2}
  \Delta\delta\geqslant\frac1{\sqrt m}\frac1{\sqrt{F_{\rm imb}}}.
\end{equation}
Here, $F_{\mathrm{imb}}$ is the Fisher information for the population imbalance measurement. It is related to the conditional probability $p(n|\delta)$ for detecting $n$ given $\delta$ as follows
\begin{equation}\label{pop}
  F_{\rm imb}=\sum_{n=0}^N\frac1{p(n|\delta)}\left(\frac{\partial p(n|\delta)}{\partial\delta}\right)^2.
\end{equation}
The above probability results from the projection of the output state onto a state with $n$ particles in one mode and $N-n$ in the other
\begin{equation}\label{prob_imb}
  p(n|\delta)=|\langle n,N-n|\hat U\ket\psi|^2.
\end{equation}
The Fisher information (\ref{pop}) through the CRLB (\ref{crlb2}) provides the maximal precision for the estimation of $\delta$ from the population imbalance measurement, whichever estimator is used.
Moreover, $F_Q\geqslant F_{\rm imb}$ always holds since the QFI sets the ultimate CRLB optimized over all the possible measurements.

Although the Fisher information from (\ref{pop}) is ``the best you can get'' from the population imbalance measurement, reaching the bound (\ref{crlb2}) requires the knowledge of the full probability (\ref{prob_imb}). 
This renders the Fisher information approach impractical in most of the cases because in order to know (\ref{prob_imb}) one must go through a laborious calibration stage. 
Therefore, typically some simpler estimators, which still utilize the data acquired from the measurements of the population imbalance, are used. The simplest estimator is based on the knowledge of the
lowest moment of (\ref{prob_imb}), namely the average, which is equal to the mean of the population imbalance operator $\hat J_z$ 
\begin{equation}\label{av_imb}
  \av{n(t)}=\sum_{n=-\frac N2}^{\frac N2} \left(n-\frac N2\right)\,p(n|\delta)=\av{\hat J_z(t)}.
\end{equation}
This average can be evaluated in the Heisenberg picture, where the $\hat J_z$ reads
\begin{equation}
  \hat J_z(t)=\hat U^\dagger\hat J_z\hat U.
\end{equation}
Using the evolution operator (\ref{evo}), we obtain 
\begin{equation}\label{jzt}
  \hat J_z(t)=u_x(t)\hat J_x+u_y(t)\hat J_y+u_z(t)\hat J_z.
\end{equation}
The three time-dependent coefficients are
\begin{subequations}\label{coeff_u}
  \begin{eqnarray}
    &&u_x(t)=\frac{\epsilon  \left[\cos \left(\varphi\sqrt{\epsilon ^2+1}\right)-1\right]  }{\epsilon ^2+1}\\ 
    &&u_y(t)=-\frac{\sin\left(\varphi\sqrt{\epsilon ^2+1}\right)  }{\sqrt{\epsilon ^2+1}}\\
    &&u_z(t)=\frac{\cos \left(\varphi\sqrt{\epsilon ^2+1}\right)+\epsilon ^2 }{\epsilon ^2+1}.
  \end{eqnarray}
\end{subequations}

\begin{figure}[htb!]
  \includegraphics[clip, scale=1]{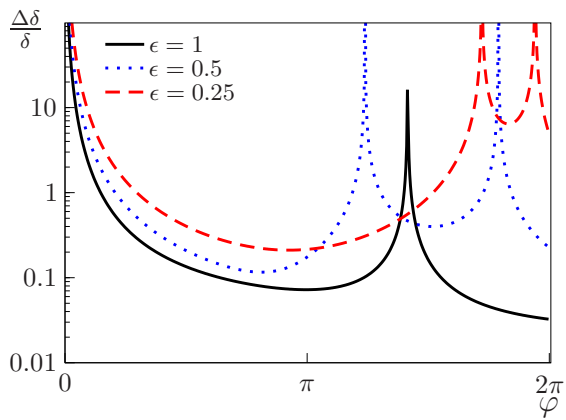}
  \caption{
    (color online) The sensitivity $\Delta \delta$ in units of $\delta$ calculated using the error propagation formula for a spin-coherent state with $N=100$. The Figure shows the
    Eq.~(\ref{ep_coh}) as a function of $\varphi$ for three different values of $\epsilon=1$ (solid black line), $\epsilon=0.5$ (dotted blue line), and $\epsilon=0.25$ (dashed red line).
  }\label{av_coh}
\end{figure}

\begin{figure}[htb!]
  \includegraphics[clip, scale=1]{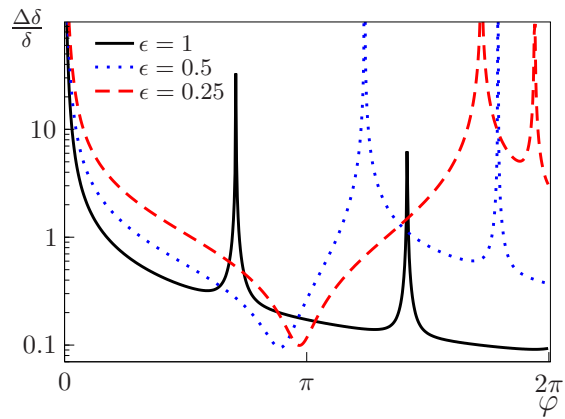}
  \caption{
    (color online) The sensitivity in units of $\delta$ calculated using the error propagation formula for a number-squeezed state of $N=100$ particles with $\xi_n^2=0.15$. The Figure shows the
    Eq.~(\ref{err_pr}) as a function of $\varphi$ for three different values of $\epsilon=1$ (solid black line), $\epsilon=0.5$ (dotted blue line), and $\epsilon=0.25$ (dashed red line).
  }\label{av_ns}
\end{figure}

\begin{figure}[htb!]
  \includegraphics[clip, scale=1]{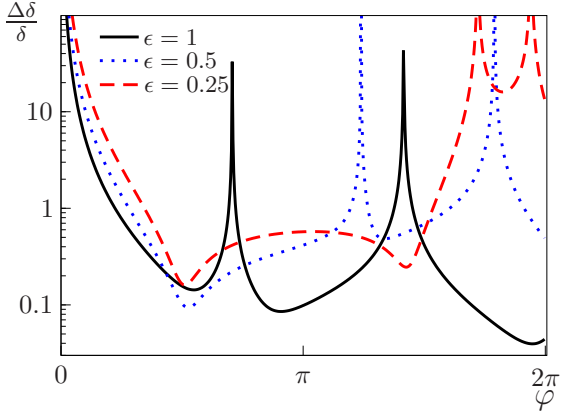}
  \caption{
    (color online) The sensitivity $\Delta \delta$ in units of $\delta$ calculated using the error propagation formula for a phase-squeezed state of $N=100$ particles with $\xi_\phi^2=0.15$. 
    The Figure shows Eq.~(\ref{err_pr}) as a function of $\varphi$ for three different values of $\epsilon=1$ (solid black line), $\epsilon=0.5$ (dotted blue line), and $\epsilon=0.25$ (dashed red line).
  }\label{av_phs}
\end{figure}

The scheme of the estimation from the average population imbalance is presented in \cite{chwedenczuk2010rabi,chwedenczuk2011phase,chwed_njp}. 
First, we assume that the function (\ref{av_imb}) is known with $\delta$ being a free parameter. In the experiment, 
this function is obtained in the calibration process. Then, the population imbalance is measured $m$ times at time $t$. 
According to the central limit theorem, if $m\gg1$ the averaged outcomes are distributed
with a Gaussian probability around the true mean value. This probability, together with the experimental outcomes, is used to construct the likelihood function $\mathcal{L}(\delta)$. 
In the final step, $\delta$ is  assigned to the value maximizing $\mathcal L(\delta)$. Such an estimator is unbiased and its sensitivity is given by the error-propagation formula
\begin{equation}\label{ep}
  \Delta\delta\geqslant\frac1{\sqrt m}\frac{\sqrt{\av{(\Delta\hat J_z(t))^2}}}{\left|\frac{\partial\av{\hat J_z(t)}}{\partial\delta}\right|}.
\end{equation}
The average and the variance of the population imbalance operator are expressed in terms of the two lowest moments of the angular momentum operators and the coefficients $u_i$. 
Combining Equations (\ref{jzt}), (\ref{coeff_u}), and (\ref{ep}), we obtain the bound for the sensitivity in units of $\delta$ 
\begin{equation}
  \frac{\Delta\delta}\delta\geqslant\frac1{\sqrt m}\frac{\sqrt{u^2_x(t)N\frac{\av{(\Delta\hat J_x)^2}}{\av{\hat J_x}^2}+u^2_y(t)\xi_\phi^2+u^2_z(t)\xi_n^2}}
       {\sqrt N\delta\left|\frac{\partial u_x(t)}{\partial\delta}\right|}.\label{err_pr}
\end{equation}
It is again a complicated function of $\epsilon$ and $\varphi$, and the input state. For a particular case of a spin-coherent state (\ref{scs}), when $\xi_n^2=\xi_\phi^2=1$, we obtain that
\begin{equation}\label{ep_coh}
  \frac{\Delta\delta}\delta\geqslant\frac1{\sqrt m}\frac{\sqrt{u^2_y(t)+u^2_z(t)}}{\sqrt N\delta\left|\frac{\partial u_x(t)}{\partial\delta}\right|}.
\end{equation}
We plot this result in Fig.~\ref{av_coh} 
as a function of $\varphi$ for the same three values of $\epsilon$ as in Fig.~\ref{qfi_coh}. We observe similar behavior as in the case of the ultimate bound discussed in Section \ref{sec_qfi}. 
For each $\epsilon$,
the sensitivity reveals some oscillatory features and the values of $\Delta\delta/\delta$ are similar to those in Fig.~\ref{qfi_coh}. To complete the comparison, in Figures \ref{av_ns} and \ref{av_phs},
we plot the expression (\ref{err_pr}) for the number-squeezed state ($\xi_n^2=0.15$) and the phase-squeezed state ($\xi_\phi^2=0.15$). Again, we observe the typical oscillatory behavior and quite similar
values of the sensitivity.

In order to gain a better insight into the precision that can be achieved from Eq.~(\ref{err_pr}), we again consider the $\epsilon\ll1$ and $\varphi\simeq1$ case. In this limit, we obtain that
\begin{equation}\label{eq:rb}
  \frac{\Delta\delta}\delta\geqslant\frac1{\sqrt m}\frac1{\sqrt N}\frac{1}{\epsilon}\frac{\sqrt{\xi^2_\phi\, \sin^2\varphi +\xi^2_n\cos^2\varphi}}{|\cos\varphi-1|}.
\end{equation}
To draw a parallel with the results from Section \ref{sec_qfi}, we first consider the case  $\varphi=\pi$ which simplifies Eq.~(\ref{eq:rb}) to
\begin{equation}\label{ep_mzi}
  \frac{\Delta\delta}\delta\geqslant\frac1{\sqrt m}\frac1{\sqrt N}\frac{1}{2\epsilon}\xi_n.
\end{equation}
This expression resembles the sensitivity of the estimation from the average population imbalance with the MZI
\begin{equation}
  \frac{\Delta\delta_{\rm MZI}}\delta\geqslant\frac1{\sqrt m}\frac1{\sqrt N}\frac{1}{\theta}\xi_n,
\end{equation}
just as Eq.~(\ref{like_mzi}) resembles the ultimate bound of the MZI.
Again, the ratio of those two is equal to $\varphi/2$. Nevertheless, the precision (\ref{ep_mzi}) improves below the SNL, if the interferometer
is fed with a squeezed state with $\xi_n^2<1$. 

The other distinguished instant of time is when $\phi=\frac\pi2$. At this point, Eq.~(\ref{eq:rb}) transforms into
\begin{equation}\label{ep_int}
  \frac{\Delta\delta}\delta\geqslant\frac1{\sqrt m}\frac1{\sqrt N}\frac{1}{\epsilon}\xi_\phi.
\end{equation}
Again, there is a close analogy between this expression and the bound (\ref{like_int}). As in the case of (\ref{like_int}), the sensitivity (\ref{ep_int}) drops below the SNL, if the input state
is phase-squeezed ($\xi_\phi^2<1$). Note, however, that the presence of $\epsilon$ in the denominator deteriorates the precision. 

\begin{figure}[htb!]
  \includegraphics[clip, scale=1]{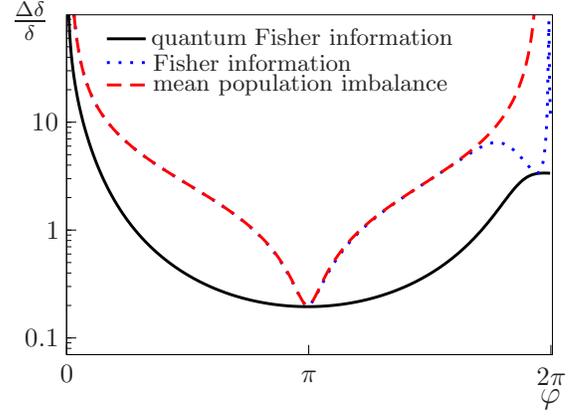}
  \caption{
    (color online) Comparison of the three bounds for the estimation precision. The solid black line is the QFI from Eq.~(\ref{qfi}). The dotted blue line is obtained 
    from the full population imbalance probability (\ref{crlb2}). The dashed red line is the error propagation formula for the estimation from the lowest moment of the population
    imbalance probability. The parameters are $\epsilon=0.1$, $\xi_n^2=0.15$, and $N=100$.
  }\label{comp}
\end{figure}

At long times, when $\varphi\gg1$, and when $\epsilon$ is small, the formula (\ref{eq:rb}) takes an appealing form
\begin{equation}
  \frac{\Delta\delta}\delta\geqslant\frac1{\sqrt m}\frac1{\sqrt N}\frac{1}{\theta\epsilon^2}\sqrt{\xi_\phi^2+\xi_n^2\cot^2\varphi}. 
\end{equation}
Again, this expression scales inversely with time. At times such that $\cot^2\varphi=0$, this sensitivity, analogically to the short-time expression (\ref{ep_int}), breaks the SNL with phase-squeezed
states. Nevertheless, the improvement from the particle entanglement might be eclipsed by the presence of $\epsilon^2$ in the denominator.

Finally, we compare the ultimate sensitivity (\ref{qfi}) with the Fisher information (\ref{crlb2}) and the error propagation formula (\ref{err_pr}). 
In Fig.~\ref{comp}, we plot the sensitivity in units of $\delta$ with $\epsilon=0.1$ as a function of $\varphi$ for a spin squeezed state with $\xi_n^2=0.15$. We observe that the simple estimation
from the average population imbalance gives the sensitivity almost as good as the Fisher information. Moreover, for $\varphi=\pi$, when Equations (\ref{like_mzi}) and (\ref{ep_mzi}) hold,
all the three methods give the same precision. This result can be explained as follows. At this time point, the sensitivities (\ref{like_mzi}) and (\ref{ep_mzi}) resemble the precision of the MZI. 
For this interferometer, the Fisher information for the population imbalance probability (\ref{crlb2}) saturates the QFI for all states (\ref{state}) with real coefficients $C_n$ \cite{optimal,lang}. 
A spin-squeezed
ground state of the Hamiltonian (\ref{bh}) satisfies this condition, therefore, expressions (\ref{like_mzi}) and (\ref{crlb2}) must coincide. 
On the other hand, such a state is Gaussian, meaning that it is characterized by the two lowest correlation functions. Not surprisingly, in such a case the sensitivity, which depends on these
two moments (\ref{ep}), is as powerful as the estimation from the full probability (\ref{crlb2}). 
To summarize, at $\varphi=\pi$ the simple estimation protocol from the average population imbalance is optimal, i.e., it saturates the ultimate bound of the QFI.

\section{Conclusions}
\label{sec_con}

We performed a systematic study of an atom interferometer which can be implemented in a double-well potential. 
This interferometer combines the phase imprint and the mode mixing at the same time. We derived the ultimate bounds for the precision of the parameter estimation and shown that
these bounds improve from the particle entanglement of the spin-squeezed states. 
Importantly, for such an interferometer the estimation from the average population imbalance gives the sensitivity which closely resembles the expression obtained for the MZI. 
Finally, we have shown that this estimation method can be optimal at the half of the period of the Josephson oscillation.
Such an oscillation-assisted interferometer, similarly to the Mach-Zehnder interferometer, can benefit from the time scaling of the sensitivity. However, in every limiting case, the precision of
the interferometer suffers from the loss of the signal, represented by the presence of $\epsilon$ in the denominator. 

Our work shows that a simple evolution operator (\ref{evo}) allows for an astonishing variety of interferometric scenarios, though
the presented analysis is not general. We restricted our calculations only to pure states and assumed that the interactions are fully suppressed during the interferometric sequence. 
We did not take into account the impact of decoherence \cite{dobrz, huelga,szankowski}. In any realistic application, the above theory should be extended to include those effects.
Nevertheless, our results serve for two purposes. First, the idealized model tells what are the ultimate bounds for the precision of the parameter estimation. Second, these findings
provide a simple theoretical background for a further analysis. 

This work was supported by the National Science Center grant no. DEC-2011/03/D/ST2/00200.

%


\end{document}